\documentclass[aps,prb,twocolumn,superscriptaddress,showpacs]{revtex4}

\usepackage{graphicx}
\usepackage{amsmath}
\usepackage{amssymb}

\begin{document}

\title{Magnetic focusing in normal-superconductor hybrid systems: 
        a semiclassical analysis}

\author{P. Rakyta}
\affiliation{Department of Physics of Complex Systems, E{\"o}tv{\"o}s
University, H-1117 Budapest, P\'azm\'any P{\'e}ter s{\'e}t\'any 1/A,
Hungary}
\author{A. Korm\'anyos}
\affiliation{Department of Physics, Lancaster University, Lancaster,
LA1 4YB, UK}

\author{Z. Kaufmann}
\author{J. Cserti}
\affiliation{Department of Physics of Complex Systems, E{\"o}tv{\"o}s
University, H-1117 Budapest, P\'azm\'any P{\'e}ter s{\'e}t\'any 1/A,
Hungary}

\begin{abstract}

 We study a transverse electron-hole focusing effect  in a normal-superconductor 
system. The spectrum of the quasiparticles is calculated both quantum mechanically 
and in semiclassical approximation, showing an excellent agreement. A semiclassical
conductance formula is derived which takes into account the effect of electron-like
as well as hole-like quasiparticles. At low magnetic field 
the semiclassical conductance shows characteristic oscillations due to the 
Andreev reflection, while at higher fields it goes to zero. These findigs are in line
with the results of previous quantum calculations and with the expectations based on 
the classical dynamics of the quasiparticles.  

\end{abstract}


\maketitle

\section{Introduction}\label{intro:ref}

Investigation of electron-transport properties of normal-superconductor (NS)
hybrid nanostructures have attracted considerable interest both
experimentally~\cite{experiments,Benistant-1:cikk,Benistant-2:cikk,Eroms_Ulrich:cikk} and theoretically~\cite{Hoppe_Ulrich:cikk,Giazotto_Ulrich:cikk,Fytas:cikk,Fagas:cikk,Chtchelkatchev_Burmistrov:cikk} in recent years.
In NS hybrid systems a crucial physical phenomenon is the Andreev reflection\cite{Andreev}
whereby  an electron incident on a superconductor-normal interface is 
(partially) retroreflected as a hole into the normal 
 conductor and  a Cooper pair is created in the
superconductor. 
The first direct experimental observation of 
the peculiar property of the Andreev reflection, i.e. 
that all the velocity components are reversed was 
achieved by Benistant et al.\cite{Benistant-1:cikk,Benistant-2:cikk}
using the versatile tool of  
transverse electron focusing (TEF)\cite{Tsoi-review:cikk}. 
The experimental and theoretical investigation of the two-dimensional
electron gas using the TEF technique has been pioneered by van Houten 
et al.~\cite{Houten_Carlo1:cikk} ( see also a 
recent review\cite{Houten_Carlo2:cikk} discussing 
these experiments in terms of coherent electron optics). 

Recently, Hoppe et al.~\cite{Hoppe_Ulrich:cikk} studied theoretically the 
interplay of the  Andreev reflection and cyclotron motion of quasiparticles 
at the interface of semi-infinite superconductor and normal 
regions in a strong  magnetic field parallel with the interface. 
They found that similarly to the normal  quantum-Hall systems, 
edge states are formed which propagate along the NS interface
but  these "Andreev" edge states consist of coherent superposition of 
electron and hole excitations. Therefore they are a new type of current carrying 
edge states which are induced by the superconducting pair potential. 
The authors of Ref.~\onlinecite{Hoppe_Ulrich:cikk} also showed that
a semiclassical  approximation can give  
a good agreement for the energy dispersion of the Andreev edge states
with the exact results obtained by solving the 
Bogoliubov-de Gennes equation~\cite{BdG-eq:konyv} (BdG).  
A clear experimental evidence for the electron and hole transport in edge states
have been reported by Eroms et al.~\cite{Eroms_Ulrich:cikk} . 

In a disk geometry, it was shown in Refs.~\cite{sajat-1:cikk,sajat-2:cikk}
that such edge states can exist both in the presence but also in the 
absence of any magnetic field and  that the
bound state energies  calculated semiclassically 
agree very well with  the results obtained from the BdG equation.

Giazotto et al.~\cite{Giazotto_Ulrich:cikk} have extended the 
study presented in Ref.~\onlinecite{Hoppe_Ulrich:cikk} by considering 
the effect of  the Zeeman splitting and of the 
diamagnetic screening currents in the superconductor on the 
Andreev edge states. 
Very recently, Fytas et al.~\cite{Fytas:cikk} studied the magnetic field 
dependence of the transport through a systems consisting of a
normal billiard and a superconducting island, while 
Chtchelkatchev and Burmistrov have also 
considered the role of the surface roughness 
in NS junctions~\cite{Chtchelkatchev_Burmistrov:cikk}.

\begin{figure}[htb]
\includegraphics[scale=0.4]{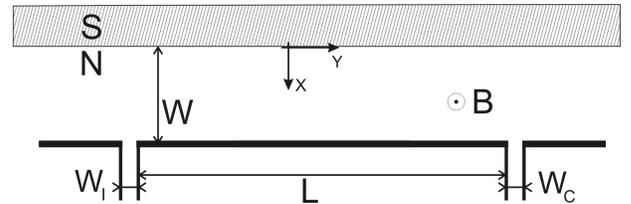}
\caption{\label{geo:fig}  
The hybrid NS nanostructure  that we investigate. It consists of 
an infinitely long two dimensional ballistic normal conductor 
 of width $W$ 
coupled to a semi-infinite spin-singlet superconductor region. 
The conductance is measured between two normal conducting 
quantum point contacts: an injector (of width $W_i$) and a collector 
(of width $W_c$ and at distance $L$ from the injector). 
The magnetic field $B$ is applied perpendicular to the system 
(in our calculation $B>0$ corresponds to a field pointing out of   
the plane of the system).}
\end{figure}

In this work we consider the NS hybrid system depicted in
Fig.~\ref{geo:fig}. 
It is similar to the experimental setup of Ref.~\onlinecite{Benistant-1:cikk} but
in our system the normal conducting region is a two-dimensional electron gas.  
It is assumed that the quasiparticle transport in the waveguide 
is ballistic and that the waveguide can  
act as drain which absorb any quasiparticles exiting to the left of the 
injector   or  to the right of the collector.

First, we calculate the eigenstate of such a system where the quantum point 
contacts are not present, ie we consider a normal waveguide in contact 
with a superconductor.   We show how the interplay of the 
lateral confinement brought about by the finite width of 
the lead, the applied
magnetic field and the proximity effect gives rise to a rich  physics in this system.
We calculate the eigenstates of the system by solving the BdG
equation and then these exact quantum results are compared with results 
obtained from  semiclassical calculations. As we shall show below the agreement 
is excellent. We note that the semiclassical approximation we used is applicable  
in a wider parameter range than the one used in 
Refs.~\onlinecite{Hoppe_Ulrich:cikk,Giazotto_Ulrich:cikk} for a similar system.

Having obtained the exact eigenstates and their semiclassical approximations, 
we then turn to the calculation of conductance for the case when 
the two quantum point contacts, as depicted in Fig.~\ref{geo:fig}, are present. 
The conductance between the injector and the collector is  determined 
by adopting the method used by van Houten et al.~\cite{Houten_Carlo1:cikk}
to describe the TEF in a two dimensional electron gas. 
In our system  however, we have to take into account  the dynamics of both 
the electron and of the hole quasiparticles.  Thus, our results on the conductance 
 can be considered  as a generalisation  of the corresponding 
semiclassical calculations of  Ref.~\onlinecite{Houten_Carlo1:cikk}.

Recently, the same  NS system has been studied
numerically~\cite{sajat-3:cikk} using a Green's function
technique~\cite{sanvito}. The influence of the  underlying classical dynamics
on the conductance has been, however,  discussed only qualitatively. 
Our rigorous semiclassical treatment gives a quantitative analysis of the 
dependence of the transport on the classical dynamics of quasiparticles.

The rest of the paper is organised as follows. 
In Section~\ref{quantum:sec} we present the exact quantum calculations
based on the BdG equation.  Then in Section~\ref{semiclassics:sec} we 
discuss the results of the semiclassical approximations of the 
exact quantum calculations. 

In Section~\ref{disp-rel:sec}
we compare the results of the quantum and of the semiclassical  calculations 
and give the physical interpretation of the results. Section~\ref{coh_foc:sec}
is devoted to the semiclassical calculation of the conductance between the 
injector and the collector which is  the central result of our paper. 
Finally, in Section~\ref{conl:sec} we come to  our conclusions.

\section{Quantum calculation}\label{quantum:sec}

In this section we consider a system consisting of a normal conducting waveguide 
of width $W$ in contact with a semi-infinite superconducting region. We 
derive a secular equation  solutions of which give
the energies of the bound states of the  system. 

The eigenstates and eigenenergies can be obtained from  the BdG equation: 
\begin{equation}
\left(\begin{array}{cc}
H_0  & \Delta \\ 
\Delta^* & -H_0^* 
\end{array}   
 \right) \mathbf{\Psi} (x,y)= E \,  \mathbf{\Psi}(x,y),
\end{equation}
where  $\mathbf{\Psi}$ is a two-component wave function and $H_0 = {\left(
{\bf p} - e{\bf A}\right)}^2/(2m)+ V-E_{\rm F}$ is the single-electron Hamiltonian
 (for simplicity, we assume that the  effective  mass $m$ and the Fermi 
energy $E_{\rm F}$  is the same  in the N and S regions\cite{note-1}).
The excitation energy $E$ is measured relative to $E_{\rm F}$. 
Scattering at the NS interface is modeled by an external potential
$V(x)=U_0\delta(x)$.   
Hard wall boundary condition  is imposed 
at the wall of the wave guide which is 
not adjacent to the superconductor, i.e. $\mathbf{\Psi}(x=W,y)=0$.
The bound state energies are 
the positive eigenvalues $0 < E < \Delta$ of the BdG equation\cite{BdG-eq:konyv}. 
The superconducting pair potential $\Delta$ can be approximated 
by a step function $\Delta({\bf r})=\Delta_0\Theta(-x)$ 
without changing the results in any qualitative way\cite{ref:andreev_billiards}.
Owing to the translational symmetry along the $y$ direction it is
convenient to chose the Landau gauge for the vector potential: 
${\bf A}({\bf r}) = B\, {(0,x,0)}^T$ 
(here $T$ denotes the transpose of a vector). 
Thus, the wave function 
$\mathbf{\Psi}(x,y)= 
{( \mathbf{\Psi}_{\rm e}(x,y), \mathbf{\Psi}_{\rm h}(x,y))}^T$  
can be separated and in the N region it reads  
\begin{equation}
\mathbf{\Psi^{(N)}}(x,y)=\begin{pmatrix}
A_e \, \Phi_{\rm  e}^{(N)}(x) \\ 
A_h \, \Phi_{\rm h}^{(N)}(x) 
\end{pmatrix}e^{ik y}, 
\label{wfn_N:eq}
\end{equation}
where $k$ is the wave number along the $y$ direction and 
the amplitudes $A_{e,h}$ will be determined from 
the boundary conditions given below.  Substituting 
$\mathbf{\Psi^{(N)}}(x,y)$ into the BdG equation we find that 
the  function $\Phi^{N}_{e}(x)$ 
satisfies the following  one-dimensional Schr\"odinger equation:
\begin{subequations}
\begin{equation}
\frac{d^2\Phi_{\rm e}^{(N)}(\xi)}{d\xi^2}-
\left(\frac{1}{4}\xi^2 + a\right)\Phi_{\rm e}^{(N)}(\xi) = 0, 
\label{eq:1DSchrodinger}
\end{equation}
where
\begin{eqnarray}\label{eq:torzs-parameterek2}
\xi=\sqrt{2}\left(\frac{x}{l}-\text{sign}(eB) k l \right), \,\, 
a=-\left(\frac{E}{\hbar\omega_{\rm c}}+\frac{\nu_0}{2}\right).   
\end{eqnarray}
\end{subequations}
Here $l=\sqrt{\hbar/{\left|eB\right|}}$ is the magnetic length, 
$\omega_{\rm c}=\left|eB\right|/m$ is the cyclotron
frequency, and 
$\nu_0=2E_{\rm F}/(\hbar\omega_{\rm c})$ is the filling factor.
Equation (\ref{eq:1DSchrodinger}) is a parabolic cylinder
differential equation~\cite{Abramowitz} 
and its solution can be expressed in terms of
the Whittaker functions $U(a,\xi)$ and $V(a,\xi)$:
\begin{equation}
\Phi_{\rm e}^{(N)}(x)= U(a,\xi) 
- \frac{ U(a,\xi_W)}{ V(a,\xi_W)}\, V(a,\xi),
\label{eq:torzs-psieN}
\end{equation}
where $\xi_W = \sqrt{2}\left(W/l - \text{sign}(eB) k l\right)$.
Note that the function $\Phi_{\rm e}^{(N)}(x)$ satisfies 
Dirichlet boundary condition at $x=W$.  
From the BdG equation it follows 
that for the hole component  
$\Phi_{\rm h}^{(N)}(x)$ of the wave function the symmetry relation  
\begin{equation}
\Phi_{\rm h}^{(N)}(B,E,x)=\Phi_{\rm e}^{(N)}(-B,-E,x)   
\label{eq:torzs-psihN}
\end{equation}
holds.

The magnetic field is assumed to be screened from the superconducting region,
hence   the vector potential is taken to be zero
(for the case  of  finite magnetic penetration length see e.g. 
Ref.~\onlinecite{Giazotto_Ulrich:cikk}).  
Therefore, in the S region the wave function ansatz with eigenenergy $E$ 
can be written as~\cite{sajat-1:cikk} 
\begin{subequations}
\begin{eqnarray}
\!\!\!\!\mathbf{\Psi^{(S)}}(x,y) &=&\!\!\!\!
\left[C_e \!\! \begin{pmatrix} \gamma_e \\ 1
\end{pmatrix}\!\!\Phi_e^{(S)}(x) 
\! + \! C_h \!\!\begin{pmatrix} \gamma_h \\ 1
\end{pmatrix}\!\!\Phi_h^{(S)}(x)\right]\!\!e^{iky},  
\label{eq:torzs-psiS} 
\end{eqnarray}
where 
\begin{eqnarray}
\Phi_{e,h}^{(S)}(x) &=& 
e^{\pm i q_{e,h} x }, \\
q_{e,h} &=& 
k_{\rm F} \sqrt{1-\frac{k^2}{k^2_{\rm F}}\mp i\eta}, 
\end{eqnarray}%
\label{Phi_e_h:eq}%
\end{subequations}%
$\eta=\sqrt{\Delta_0^2-E^2}/E_{\rm F}$, 
$\gamma_{e,h}=\frac{\Delta_0}{E\pm i\sqrt{\Delta_0^2-E^2}}$, and
$k_{\rm F}= \sqrt{2m E_{\rm F}}/\hbar$ is the Fermi wave number 
(the upper/lower signs in the expressions correspond to the
electron/hole component). 
Note that in the S region the wave function $\mathbf{\Psi}^{(S)}(x,y)$ 
goes to zero for $x\to -\infty$.

The four coefficients $A_{e,h},  C_{e,h}$ in
Eqs.~(\ref{wfn_N:eq}) and (\ref{eq:torzs-psiS}) are determined by the
the boundary conditions
at the NS interface\cite{sajat-1:cikk,sajat-2:cikk}:
\begin{equation}
\begin{split}
\big.\mathbf{\Psi^{(N)}}\big|_{x=0} &= 
\big.\mathbf{\Psi^{(S)}}\big|_{x=0},  \\
\left.\frac{d}{dx}\left[\mathbf{\Psi^{(N)}}
-\mathbf{\Psi^{(S)}}\right]\right|_{x=0} &= 
\frac{2m}{\hbar^2}U_0\big.\mathbf{\Psi^{(N)}}\big|_{x=0}, 
\end{split}	 
\label{eq:torzs-hatarfeltetel0}
\end{equation}
for all $y$ and $k$. 
The matching conditions shown in Eq.~(\ref{eq:torzs-hatarfeltetel0})
 yield  a secular equation for the eigenvalues $E$ 
as  a function of the  wave number $k$. 
Using  the
symmetry relations between the electronic and hole-like component of
the BdG eigenspinor given by Eq.~(\ref{eq:torzs-psihN}), the secular equation can be reduced 
to~\cite{sajat-1:cikk,sajat-2:cikk}
\begin{subequations}
\begin{equation} 
\rm{Im} \left \{\gamma_e D_e(E,B) \, D_h(E,B) \right \}=0,
\label{DNS}
\end{equation}
where the $ 2\times 2$ determinants $D_e$ and $D_h$ are given by 
\begin{eqnarray}
D_e(E,B)  &=&  
\left| \begin{array}{cc}
 \Phi_e^{(N)} & \Phi_e^{(S)}  \\
{\left[\Phi_e^{(N)}\right]}^\prime 
& Z\Phi_e^{(S)} + {\left[\Phi_e^{(S)}\right]}^\prime
\end{array}   \right|  , \\ 
D_h(E,B) &=& D_e(-E,-B) .
\label{De}
\end{eqnarray}%
\label{sec-exact:eq}%
\end{subequations}%
Here  $Z= \left(2m/\hbar^2\right) \, U_0$ is
the normalised barrier strength, and the prime denotes the derivative
with respect to $x$. All functions are evaluated at the NS interface ie. at $x=0$. 
The secular equation derived above is exact in the sense that the
usual Andreev approximation is not assumed~\cite{ref:andreev-approx}. 
An analogous result was found previously~\cite{sajat-1:cikk,sajat-2:cikk} 
for NS disk systems.

\section{Semiclassical approximation}\label{semiclassics:sec}

As a first step to calculate the conductance between the injector and collector 
we should solve the exact quantisation condition (\ref{sec-exact:eq}) which  
involves evaluation of parabolic cylinder functions. 
It turns out that  for certain parameter ranges this makes the actual numerical
calculations rather difficult. However, as we are going to show it in 
Section \ref{disp-rel:sec},  for the quasiparticle dispersion
relations which will be  important in the subsequent analysis, 
one can obtain  excellent approximations using semiclassical methods.
The use of  the semiclassical approximations makes  
the numerical calculations  much simpler and    gives a   
better understanding of the underlying physics. 
The semiclassical calculations are based on (i) the WKB approximation\cite{Brack:konyv} 
of the functions $\Phi_e(x)$, $\Phi_h(x)$ 
[see Eqs.~(\ref{eq:1DSchrodinger}) and (\ref{eq:torzs-psihN})] and their derivatives 
(ii) the Andreev-approximation\cite{ref:andreev-approx}. 
The approximated wave functions are then 
substituted into Eq.~(\ref{DNS}) to obtain the semiclassical quantisation conditions.  
The calculations can be carried out in a similar fashion as in 
Ref.~\onlinecite{sajat-2:cikk}, therefore in 
this section and in the next one we only summarise the main results. 
Throughout the rest of the paper we assume ideal NS interface, ie we set $U_0=0$.

\begin{figure*}
\centering
\includegraphics[width=14cm]{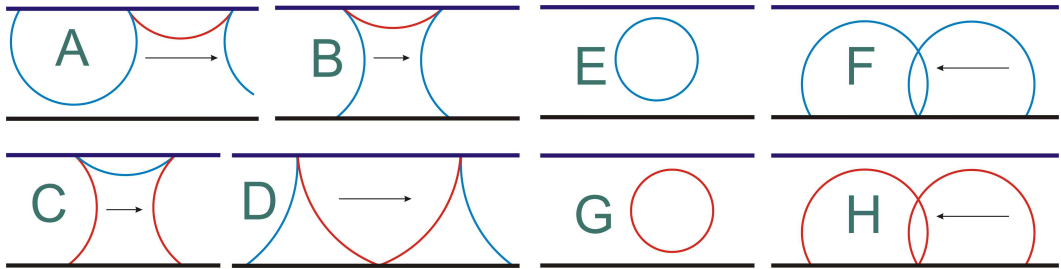}
\caption{Classification of the possible orbits.  
The blue/red lines correspond to electron/hole trajectory segments. The black 
arrows show the direction of the group velocity (cf. the slope of the dispersion
curves in Figs.~\ref{fig:palyaazonositEAG} and \ref{fig:palyaazonositFBDCH}) for
a magnetic field pointing out of the plane.}
\label{fig:palyak} 
\end{figure*}

Depending on the energy of the electrons (holes) and the applied magnetic
field, eight different types of orbits can be distinguished. These orbits,
denoted by capital letters $A$ to $H$, are  shown in Fig.~\ref{fig:palyak}. 
In the geometrical construction of the classical trajectories 
we took into account  that the chiralities of the electron-like and 
the hole-like orbits  are preserved when electron-hole conversion occurs
at the NS interface~\cite{Giazotto_Ulrich:cikk}.

Type $A$ orbits correspond to  
skipping motion of  alternating electron and  hole quasiparticles
along the NS interface. 
Neither the electron's nor the hole's trajectory  hits  the wall of the
wave guide at $x=W$. 
This type of orbit was first considered by Hoppe et al.~\cite{Hoppe_Ulrich:cikk}.  
Type $B$ and $C$ orbits are similar to type A but the either the 
electron or the hole can now reach the wall of the wave guide at $x=W$. 
In case of type $D$ orbits the  electrons and holes bounce both at the  wall 
of the wave guide and  at the  NS interface, while for  type $E$ ($G$) orbits 
the quasiparticles are moving on  cyclotron orbits.
Type $F$ ($H$) is  the familiar  edge state  
of the integer quantum Hall systems.  The electrons (holes) are moving
on skipping orbits  along the  wall of the wave guide at $x=W$. 

The semiclassical quantisation condition for  the orbits shown 
in Fig.~\ref{fig:palyak} can be written in the following  simple form: 
\begin{subequations}
\begin{equation}
N(E) = n+\mu, \qquad \text{for} \quad n\in\mathbb{Z}, 
\label{semi-quant-cond:eq}
\end{equation}
where $N(E)$ can be 
expressed in terms of the (dimensionless) classical action integrals
 $S_{e}(\tau_+^{e}, \tau_-^{e})$, $S_{h}(\tau_+^{h}, \tau_-^{h})$  
 (see Table~\ref{orbits:table}) 
 for different  types of orbits and $\mu$ is   
  the corresponding Maslov index.  
  The actions $S_{e,h}$ are given by the
following equations:
\begin{eqnarray}
S_e(\tau_+^{e},\tau_-^{e}) &=& 
2\left(\Theta_{\rm e}(\tau_+^{e}) - \Theta_{\rm e}(\tau_-^{e})\right),\\
S_h &=& S_e(-E-B), \\[2ex]
\Theta_{\rm e}(x) &=& 
 \frac{\left|eB\right|}{2 \pi\hbar}\int\limits_{}\sqrt{R_{\rm e}^2 
-\left(x-X\right)^2}\;{\rm d}x 
\nonumber \\
&=& \frac{E+E_{\rm  F}}{2 \pi\hbar\omega_{\rm c}}
\left[\arcsin\frac{x-X}{R_{\rm e}} \right. 
\label{phase_int:eq}
\nonumber \\
&& + \left. 
\frac{1}{2}\, \sin\left(2\arcsin\frac{x-X}{R_{\rm e}}\right)\right], \\
\Theta_{\rm h} &=& \Theta_{\rm e}(-E,-B). 
\end{eqnarray}
Here   the cyclotron radii  $R_{\rm e, h}$ 
and the  classical turning points $\tau^{e,h}_\pm$ 
for electrons and  holes 
are given by  
\begin{eqnarray}
R_{\rm e,h} &=& R_c \sqrt{1 \pm E/E_{\rm F}}, \quad R_c = k_{\rm F}l^2, 
\label{R_e:eq}\\ 
\tau^e_+ &=& \min\left\{W,X + R_e\right\},
\label{turnpoints-e:eq1} \\
 \tau^e_- &=& \max\left\{0,X - R_e\right\},
 \label{turnpoints-e:eq2} \\
\tau^h_\pm &=& \tau^e_\pm(-B,-E),  
\label{turnpoints-h:eq}
\end{eqnarray}
\label{semiS:eq}
\end{subequations}
where $X={\rm sign} (e B) k l^2$ is the guiding
center coordinate. Finally,  
contributions to the Maslov index of a given type of orbit come from 
the collisions with the wall of the waveguide, from the collisions 
with the superconductor (Andreev reflections), and from the caustics of the 
electron (hole) segments of the orbit (see Table.\ref{orbits:table}). 
\begin{table*}[htb]
\begin{tabular}{|c|c|c|c|}
\hline
type of orbit  & $N(E)$ & $\mu$  & conditions for orbits\\
\hline
$A$ & $S_e(\tau_+^{e},\tau_-^{e}) - S_h(\tau_+^{h},\tau_-^{h})$ 
& $ \frac{1}{\pi}\arccos\left(\frac{E}{\Delta}\right)$ 
& $\tau_+^{e},\tau_+^{h}<W, \quad\tau_-^{e},\tau_-^{h}=0$\\
$B$ & $S_e(\tau_+^{e},\tau_-^{e}) - S_h(\tau_+^{h},\tau_-^{h})$ 
& $-\frac{3}{4}+ \frac{1}{\pi}\arccos\left(\frac{E}{\Delta}\right)$ 
& $\tau_+^{e}=W, \quad\tau_+^{h}<W, \quad\tau_-^{e},\tau_-^{h}=0$\\
$C$ & $S_e(\tau_+^{e},\tau_-^{e}) - S_h(\tau_+^{h},\tau_-^{h})$ 
& $-\frac{1}{4}+ \frac{1}{\pi}\arccos\left(\frac{E}{\Delta}\right)$ 
& $\tau_+^{e}<W, \quad\tau_+^{h}=W, \quad\tau_-^{e},\tau_-^{h}=0$\\
$D$ & $S_e(\tau_+^{e},\tau_-^{e}) - S_h(\tau_+^{h},\tau_-^{h})$ 
& $ \frac{1}{\pi}\arccos\left(\frac{E}{\Delta}\right)$ 
& $\tau_+^{e},\tau_+^{h}=W, \quad\tau_-^{e},\tau_-^{h}=0$\\
$E$ & $S_e(\tau_+^{e},\tau_-^{e}))$ 
& $\frac{1}{2}$ & $\tau_+^{e}<W, \quad\tau_-^{e}>0$\\
$F$ & $S_e(\tau_+^{e},\tau_-^{e})$ 
& $-\frac{1}{4}$ & $\tau_+^{e}=W, \quad\tau_-^{e}>0$\\
$G$ & $-S_h(\tau_+^{h},\tau_-^{h})$  
& $-\frac{1}{2}$ & $\tau_+^{h}<W, \quad\tau_-^{h}>0$\\
$H$ & $-S_h(\tau_+^{h},\tau_-^{h})$ 
& $-\frac{3}{4}$ & $\tau_+^{h}=W, \quad\tau_-^{h}>0$\\
\hline	
\end{tabular}
\caption{$N(E)$ related to the actions defined in Eq.~(\ref{semiS:eq}), 
and the conditions for the possible orbits shown in Fig~\ref{fig:palyak}. 
\label{orbits:table}}
\end{table*}

\section{Dispersion relation and the phase diagram for NS systems}\label{disp-rel:sec}

\begin{figure}[htb]
\centering
\includegraphics[scale=0.19]{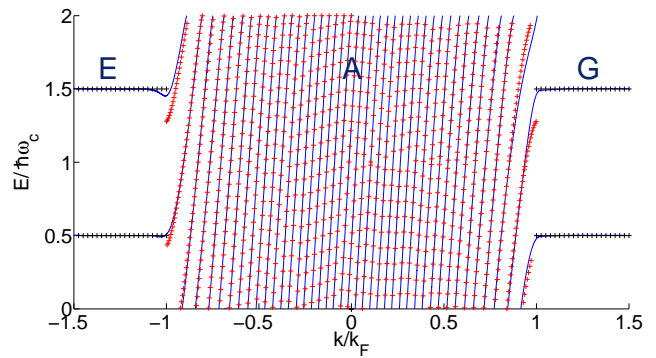} 
\caption{The energy spectrum 
obtained from Eqs.~(\ref{sec-exact:eq}) (solid/blue line) 
and (\ref{semiS:eq}) (red crosses) 
as a function of $k$. 
The parameters are $\nu_0=40$, $\Delta_0/\hbar\omega_{\rm c}=2$,  
$k_{\rm F}W=106.7$ and  $R_c/W=0.375$. 
In the semiclassical calculations only the orbits of type   
$A$, $E$ and $G$ need to be taken into account. }
\label{fig:palyaazonositEAG}
\end{figure}

\begin{figure}[htb]
\centering
\includegraphics[scale=0.19]{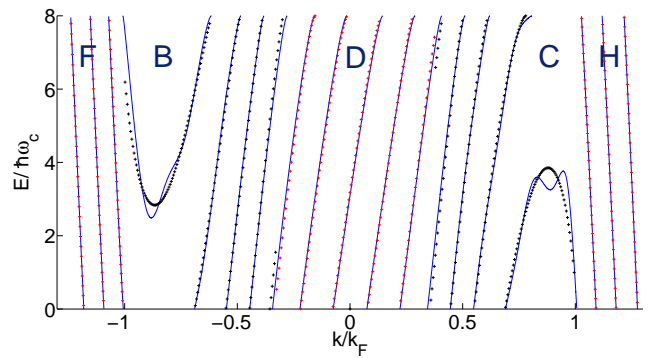} 
\caption{The same as in Fig.~\ref{fig:palyaazonositEAG} 
with parameters $\nu_0=160$, $\Delta_0/\hbar\omega_{\rm c}=8$,  
$k_{\rm F}W=106.7$ and $R_c/W=1.5$. 
In the semiclassical calculations the orbits of type 
$B$, $C$, $D$, $F$ and $H$ are involved. } 
\label{fig:palyaazonositFBDCH}
\end{figure}

In this section we compare the numerical results obtained from the
exact and from the semiclassical calculations outlined in 
Sections \ref{quantum:sec} and \ref{semiclassics:sec}. The above 
discussed classical orbits can exist in different parameter ranges,
depending on the strength of the magnetic field, on the Fermi energy and 
on the width of the normal lead. 
It is  convenient   to use the following dimensionless parameters: 
 $\nu_0$, $\Delta_0/\hbar\omega_{\rm c}$, $k_{\rm F}W$ and  $R_c/W$. 
Figures~\ref{fig:palyaazonositEAG} and \ref{fig:palyaazonositFBDCH} 
show comparisons of  the exact quantum calculations with  the semiclassical
results for the dispersion relation of the quasiparticles. 
 In case of Fig.~\ref{fig:palyaazonositEAG}  
the magnetic field is strong so  that the cyclotron radius
is smaller than the width of the waveguide. One expects therefore that 
 type A orbits and  Landau levels  
(corresponding to orbits of type $E$ and $G$) would appear in the 
spectrum. One can see that this is exactly the case, the Landau levels
appearing as dispersionless states.  The 
agreement between the quantum and semiclassical calculations is excellent
except in the transition regime between type A and type E (G) orbits.
In case of Fig.~\ref{fig:palyaazonositFBDCH} the magnetic field is weaker 
than for Fig.~\ref{fig:palyaazonositEAG} and therefore  
 $R_c$ is now larger than  $W$. 
No Landau levels appear  and the dispersion 
relation can be well approximated semiclassically using  orbits of 
type $B$, $C$, $D$, $F$ and $H$.

\begin{figure}[htb]
\centering
\includegraphics[scale=0.4]{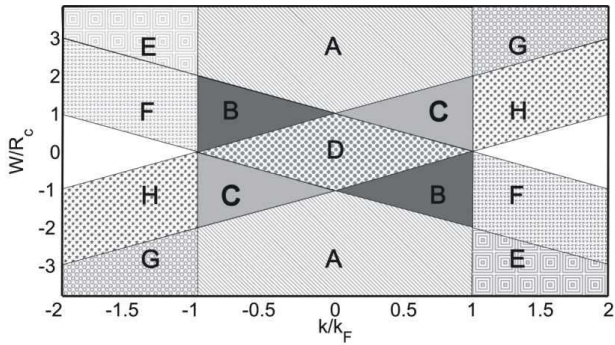} 
\caption{ 
The phase diagram of the allowed  type of orbits as a function of $R_c$  and $k$.
The white regions are classically forbidden. 
The energy of the quasiparticles is $E=0$.
}
\label{fig:phasediagram}
\end{figure}

Whether or not a given type of classical orbit is allowed for a certain set of 
parameter values depends on the  positions of 
the turning points with regard to the wall of the wave guide and to the NS interface.
The conditions for each type of orbits are summarised in Table~\ref{orbits:table}. 
The turning points  depend  [see Eqs.~(\ref{R_e:eq}) -- (\ref{turnpoints-h:eq})]
on the width of the  lead $W$, on the Fermi wavenumber $k_{\rm F}$, 
on the magnetic field (or equivalently, on the cyclotron radius $R_c$),
and on the wave number $k$. 
In an experiment the former two parameters would be fixed, 
$R_c$ could be varied  by varying the magnetic field, while
the wave number $k$ of the injected electrons would be uniformly 
distributed if there are many open channels in the quantum point contact.
For a given $W$ and $k_{\rm F}$, ie for a given experimental sample, 
 one can then draw a "phase diagram", 
which shows the allowed types of classical orbits as a function of
$R_c$ and  $k$.  An example of such a diagram is shown in Fig.~\ref{fig:phasediagram}.  
Note that the weak energy dependence of the turning
points translates into a similarly weak energy dependence of the  phase diagram.
Thus for  Fig.~\ref{fig:phasediagram} we have chosen $E=0$. 
The white regions are classically forbidden, as no classical
orbit can be realized for  these parameter values.

\section{Coherent electron-hole focusing}
\label{coh_foc:sec}

Having obtained the spectrum of the quasiparticles, we have now 
all the necessary information to  calculate  
the conductance between the injector and 
collector for the NS system  shown 
in Fig.~\ref{geo:fig}. 
For $W_i,\,\,W_c \ll \lambda_F$  the calculations in principle 
could be carried out using the exact results of Section \ref{quantum:sec}. 
However, this would involve evaluations of 
the Whittaker functions [see Eq.~(\ref{eq:torzs-psieN})] which would render the numerical 
calculations rather difficult. Therefore we calculate the conductance semiclassically, 
adopting and generalising  the method of Ref.~\onlinecite{Houten_Carlo1:cikk} 
to account for all types of  current carrying modes. Namely, the dynamics of hole type quasiparticles created by Andreev reflections also needs to be taken into consideration.  
We assume that $W_i, W_c \ll R_c$ meaning that the angular distribution of the injected 
electrons is not perturbed by the magnetic field.
 
In classical picture injected electrons having orbits of type B or 
D can contribute to the conductance, since only these orbits can reach the collector 
(note that for  type F orbit the group velocity points in the $-y$ direction).
Assuming that the wave function in the wave guide is unperturbed by the
presence of the  collector\cite{Houten_Carlo1:cikk}, the   
current at the collector is given by 
\begin{equation}
I_{\rm c}=\varepsilon 
\left(\left|\frac{\partial \mathbf{\Psi}_{\rm e}}{\partial x}(W,L)\right|^2 
- \left|\frac{\partial \mathbf{\Psi}_{\rm h}}{\partial x}(W,L)\right|^2\right), 
\label{eq:kollektoraram}
\end{equation}
where $\varepsilon $ is an yet undetermined parameter but will drop out 
when we calculate  the conductance.
This expression is the generalisation of Eq.~(16) in
Ref.~\onlinecite{Houten_Carlo1:cikk} by including the contribution of
the holes.  
In WKB approximation the wave function 
$\mathbf{\Psi^{(N)}}(x,y)=(\mathbf{\Psi}_{\rm e}(x,y), \mathbf{\Psi}_{\rm h}(x,y))^T$  
in the wave guide is the sum over all classical trajectories 
from the injector to the point $(x,y)$ of an amplitude factor times a 
phase factor. 
As in Ref.~\onlinecite{Houten_Carlo1:cikk} we transform the sum over
trajectories into a sum over  modes using saddle point
integration. 
Finally, we find 
\begin{subequations}
\begin{eqnarray}
\lefteqn{ 
\frac{\partial\mathbf{\Psi}_{\rm e}}{\partial x}(W,L) =} \nonumber\\
&& \hspace{-11mm} -2ik_{\rm F} \hspace{-3mm}
\sum_{\scriptstyle n \atop \scriptstyle K=B,D}  \hspace{-3mm} 
\sqrt{2\pi i \! \left(\!\!\frac{\partial^2 S^K_e(p_n)}
{\partial p_n^2}\!\! \right)^{-1}} A_{p_n}^K e^{ik_n^K L - i\pi} 
\cos\alpha_{n}^K,     
\label{eq:torzs-hullamfv-modusokelektron}
\end{eqnarray}
 and 
\begin{eqnarray}
\lefteqn{ 
\frac{\partial\mathbf{\Psi}_{\rm h}}{\partial x}(W,L) =} \nonumber\\
&& \hspace{-11mm} -2ik_{\rm F}  \hspace{-1mm}
\sum_{n} \sqrt{2\pi i \! \left(\!\!\frac{\partial^2 S^D_h(p_n)}
{\partial p_n^2}\!\! \right)^{-1}} A_{p_n}^D e^{ik_n^D L - i\pi} 
\cos\alpha_{n}^D.  
\label{eq:hullamfv-modusokhole} 
\end{eqnarray}%
\label{Psi_deriv:eq}%
\end{subequations}%
For simplicity, here we give the definitions of the different terms appearing 
in the above expressions only for electrons having type $B$ orbits.
For holes and  for type $D$  orbits similar expressions were derived but are not
presented here. 
The wave numbers $k_n$ of the excited modes (for a given magnetic field) 
can be obtained from 
the dispersion relation by solving the equation $E(k_n,R_c)=E$. 
The amplitudes $A^B_{p_n}$ of the modes related to type $B$ of orbits 
are given by
\begin{equation}
A^B_{p_n} = \sqrt{\frac{I_i \cos \alpha_n }{2v_{\rm F} L} \, 
\frac{d(\alpha_n)}{d^\prime(\alpha)|_{\alpha = \alpha_n}}}, 
\label{eq:amplitude}
\end{equation}
where $I_i$ is the current injected from the injector, $v_{\rm F} =
\hbar k_{\rm F}/m $ is the Fermi velocity of the quasiparticles and
the prime denotes derivation with respect to $\alpha$. Here $\alpha$,
$\alpha_n$ and $d(\alpha_n)$ are defined in the following way:
the distance between two subsequent bouncing at the wall of the wave guide 
for an electron injected at angle $\alpha$ (measured from the $y$ axis) is 
$d(\alpha)=4\sqrt{R_c^2-{\left(W-R_c \sin \alpha\right)}^2} 
- 2R_c \cos \alpha$. 
Then $p=L/d(\alpha)$ is the number of bounces  between the injector and collector. 
The angle $\alpha$ can also be expressed by $\sin \alpha= (W-X)/R_c$ 
and it can be related to the wave number  $k_n$ of the modes. 
Since for mode $k_n$ the guiding centre coordinate is $X={\rm sign}(eB) k_n l^2$, 
we have $\sin \alpha_n = W/R_c - {\rm sign}(eB) k_n/k_{\rm F}$ 
(because $R_c = k_{\rm F} l^2$), and then $p_n = L/d(\alpha_n)$.
Finally, $S^B_{e,\,(h)}(p_n)$ is related to the action of  electrons (holes)  
calculated from the injector to the collector. 
In case of type $B$ orbits the summation over $n$ includes only 
those modes for which the group velocity $\partial E(k)/\partial (\hbar k)$ 
is  positive, i.e the mode propagates from the injector 
to the collector. (Note that the group velocity is basically given by the slope of
the curves in Fig.\ref{fig:palyaazonositFBDCH}).

The conductance $G(E,R_c)$ between the injector and the collector is
$G=I_i/V_c$, where $V_c$ is the collector voltage. 
Taking into account Eq.~(\ref{eq:kollektoraram}) it 
can be calculated\cite{Houten_Carlo1:cikk} as  
\begin{equation}
V_{\rm c} = \varepsilon\left(
\frac{\left|\frac{\partial\Psi_{\rm e}}{\partial x}(W,L)\right|^2}
{G_{\rm e}} 
- \frac{\left|\frac{\partial\Psi_{\rm h}}{\partial x}(W,L)\right|^2}
{G_{\rm h}}\right).   
\label{eq:konduktanciakvantum}
\end{equation}
Here $G_e$ ($G_h$) is the conductance of the point contact of the
collector for electrons (holes). It is estimated by Eq.~(17) in 
Ref.~\onlinecite{Houten_Carlo1:cikk}. 
Since both $G_e$ and $G_h$ are proportional to 
the parameter $\varepsilon$, it drops out from $V_c$. 
Similarly, the injected current $I_i$  drops out from $G(E, R_c)$ 
since the derivatives of the wave functions in (\ref{Psi_deriv:eq})
are proportional to $\sqrt{I_i}$ through the amplitudes $A_{p_n}$.  

\begin{figure}[htb] 
\centering
\includegraphics[scale=0.4]{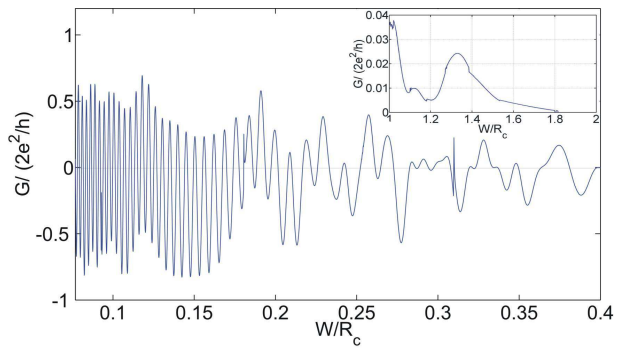} 
\caption{ 
The conductance $G$ for low  magnetic fields  at  $E=0$.  
Inset: the conductance for stronger magnetic fields. 
We used  $k_{\rm F}W=26.7$.
}
\label{fig:conductance}
\end{figure}

Equations (\ref{eq:kollektoraram}), (\ref{Psi_deriv:eq}) and 
(\ref{eq:konduktanciakvantum}) 
allow us to calculate the conductance semiclassically 
as a function of the energy of quasiparticles  and of the  magnetic field. 
This is the main result of our paper. 
An example is shown in Fig.~\ref{fig:conductance}.  
One can clearly see that   for small magnetic fields 
($W/R_c< 0.4$) the conductance is a rapidly oscillating function of
the magnetic field and can also be negative. 
The first observation is the consequence of the constructive
interference of many coherently excited modes (edge states). 
Similar effect was found 
by van Houten et al.~\cite{Houten_Carlo1:cikk,Houten_Carlo2:cikk}. 
The fact that the conductance can be negative is a consequence of  
the hole-like excitations in the hybrid NS system. 
Indeed, at low magnetic fields (large $R_c$) an  injected electron will
 undergo one or more Andreev reflections and it can happen (see type D orbits)
that a hole will arrive to the collector resulting in negative conductance.
This is the so called \emph{Andreev-drag effect}\cite{sajat-3:cikk}. 
The heights of the positive peaks are comparable with those of 
the negative ones, indicating that it can be a robust effect.   
This observation is in line with the findings of Ref.~\onlinecite{Benistant-1:cikk} 
and also with the results of Ref.~\onlinecite{sajat-3:cikk}
where an exact (numerical) quantum calculation has been performed 
for the same system.

The use of WKB approximation means that our results should be 
accurate at low magnetic fields when a large number of edge states 
are populated. Nevertheless, we find that this method, at least qualitatively,       
also describes the regime of high magnetic fields. Namely, from the 
inset of Fig.~\ref{fig:conductance} one can see that 
increasing the magnetic field the conductance decreases and
rapidly  goes to  zero for $W/R_c > 1.5$. 
This can be understood from classical considerations. 
Increasing the magnetic field the cyclotron radius decreases and at
a certain value of the field we have $2 R_c < W$  meaning that no 
injected electron can hit the superconductor and undergo Andreev
reflections. Instead, the electrons  move to the left skipping
along the wall of the wave guide (type $F$ of orbits) 
and eventually they leave the system without reaching the collector, ie the
conductance becomes zero. According to our semiclassical calculations, 
for a system with $k_{\rm F}W=26.7$ as in case of Fig.\ref{fig:conductance} 
the last current-carrying mode disappears 
when $W/R_c \gtrsim 1.8$, in broad agreement with the classical picture.

\section{Conclusions} \label{conl:sec}

In conclusion, we have studied the transverse electron-hole 
focusing effect in a normal-superconductor system  similar 
to the setup of Ref.~\onlinecite{Benistant-1:cikk}.
As a first step to determine the conductance, we calculated 
the energies of the bound states both quantum mechanically and 
in semiclassical approximation. We have shown that semiclassical methods 
well reproduce the results of the relevant quantum  calculations and
thus can help to understand the underlying physics. We have identified
those classical orbits which contribute to the conductance in our 
system and derived a semiclassical formula for the conductance.
In agreement with the quantum calculations of Ref.~\onlinecite{sajat-3:cikk}, 
for weak magnetic fields 
the  semiclassical conductance shows rapid oscillations  and the 
presence of the Andreev-drag effect. 
 For stronger magnetic fields
the conductance goes to zero which can be understand invoking the classical
dynamics of electrons at such fields. 
Our results can be considered  generalisation of 
similar works\cite{Houten_Carlo2:cikk,Houten_Carlo1:cikk} for normal 
systems since in our case the current carrying modes are comprised of   
electron-like as well as hole-like quasiparticles.

\acknowledgments

This work is supported by European Commission Contract No.~MRTN-CT-2003-504574.


\begin{thebibliography}{10}
\bibitem{ref:andreev-approx}
In Andreev approximation, making use of the fact that usually
$\Delta_0/E_{\rm F} \ll 1$ and assuming that the quasiparticles are 
incident at the NS interface almost
perpendicularly, one neglects the normal reflection. 
See, e.g., Ref.~\onlinecite{lambertreview}.
\end{thebibliography}
\end{document}